\documentclass[11pt]{article}
\usepackage{epsfig, graphics, graphicx, amsmath, amsthm, amsbsy,epsfig, fullpage}

\begin{document}

\title{The use of ideas of Information Theory for studying
``language'' and intelligence  in ants}

\author{Boris Ryabko\footnote{Siberian State University of Telecommunications and Informatics;
  Institute of Computational Technologies  of Siberian Branch
Russian Academy of Science, Novosibirsk, Russia.\ boris@ryabko.net}, 
 Zhanna Reznikova\footnote{ Institute for Animal Systematics and Ecology,
 Siberian Branch RAS and Novosibirsk State University.\ zhanna@reznikova.net}
}

\date{}

\maketitle

\abstract{
 In this review we integrate results of long term experimental study on ant "language"
 and intelligence which were fully based on fundamental ideas of Information Theory, such as
 the Shannon entropy, the Kolmogorov complexity, and the Shannon's equation connecting
 the length of a message ($l$) and its frequency $(p)$, i.e. $l = - \log p$ for rational communication systems.
This approach, new for studying biological communication systems,
enabled us to obtain the following important results on ants'
communication and intelligence: i)  to reveal "distant homing" in
ants, that is, their ability to transfer information about remote
events; ii) to estimate the rate of information transmission; iii)
to reveal that ants are able to grasp regularities and to use them
for  "compression" of information; iv) to reveal that ants are
able to transfer to each other the information about the number of
objects; v) to discover that ants can add and subtract small
numbers.

 The obtained
 results show that Information Theory is not only wonderful
 mathematical theory, but many its results may be considered as
 Nature laws.}


\section{Introduction}

Since C.Shannon published his famous paper ``A mathematical theory
of communication'' \cite{S} the fundamental role of Information
Theory has been appreciated not only in its direct applications,
but also in robotics, linguistics and biology.

    The communication systems and related cognitive skills of animals are a matter of special
    interest to ethologists, psychologists, linguists, and specialists in artificial intelligence and robotics.
    Attempts to approach the question whether highly social and intelligent species can exchange meaningful messages
    are based on a natural idea that the complexity of communication should be connected with high levels of sociality,
    cognition and cooperation in animals' societies. In the 1960s and 1970s, elegant but ambiguous experiments
    were conducted in which animals were asked to pass some pieces of information to each other.
    In Menzel's \cite{Me} experiments, a group of chimpanzees living in an enclosure searched for hidden food.
    Menzel suggested that chimpanzees possess means for transferring information about both location and properties of objects,
    but it remained unclear how they did this. In other experiments the cooperative behavior of dolphins was investigated,
    which could involve intelligent communication. To get a fish, two dolphins, separated by an opaque barrier, had to press
    the paddles in the correct order. The obtained results enabled researchers to suggest that the dolphins can co-ordinate
    the actions of each other, probably, by means of acoustic signals \cite{EB}. Despite these supportive
    experiments that demonstrated that members of highly social intelligent species really have what to "say" to each other,
    the question of existence of developed "languages" in non-human beings remained so far obscure.

       The main difficulties in the analysis of animal "languages" appear to be methodological. Many
       researchers have tried to directly decipher animal language by looking for "letters" and "words"
       and by compiling "dictionaries" \cite{Ry}, for a review see \cite{Re1A}. However, only
       two cases of natural communications have been decoded up to the present. First, one of the most
       complicated of the known natural "languages" in animals is the symbolic honey bee "Dance language".
       Discovered by K. Von Frisch \cite{Fr1,Fr2} it was later intensively studied using different methods
       including robotics and radars \cite{M,RGSRM}. The second case of a successful
       deciphering of several natural signals concerned alarm calls in vervet monkeys
       which appeared to give different alarm calls to eagles, snakes and leopards \cite{SC,CG}. Later "semantic" alarm
       calls and food calls were described for several other species \cite{BKK,HM};
       for a detailed review see \cite{Re1A}.
        The fact that researchers have managed to compile such
       "dictionaries" for a few species only, appears to indicate not that other animals lack "languages",
       but that adequate methods are lacking. In both cases of communications that became partly accessible
       for investigators, expressive and distinctive signals correspond to repeatable and frequently occurring
       situations in the context of animals' life, and thus can serve as "keys" for decoding their signals.

       The problem of cracking animals' codes have become especially attractive since the great "linguistic" potential
       was discovered in several highly social and intelligent species by means of intermediary artificial languages.
       Being applied to apes, dolphins and grey parrots, this method has revealed astonishing mental skills in the
       subjects \cite{GG,HAEH,SST,Pe}.
       It is become possible to demonstrate that animals are capable not only of decision making and using the
       experience gained in new situations, but also of using simple grammatical rules, visual symbols and
       number-related skills. For example, language-trained chimpanzees were found to be able to add and
       subtract small numbers \cite{BH}  - the ability that was not available for discovery without the use
       of intermediary languages.

     However, it is important to note that this way to communicate with animals is based on adopted human
     languages. Yet surprisingly little is known about natural communication systems of those species that were involved
     in language-training experiments. Explorers of animal language behavior thus have met a complex problem of resolving
     the contradiction between their knowledge about significant "linguistic" and cognitive potential in some species and
     limitations in cracking their natural codes \cite{MH}.

     We have suggested a principally
     new experimental paradigm based on concepts of Information Theory \cite{rr0,RR1,RR96}.
     The main point of our approach is not to decipher signals but to investigate the very process of information
     transmission by measuring time duration which the animals spend on transmitting messages of definite lengths and
     complexities. Being applied to studying communication in highly social ant species, this approach enabled us to
     reveal basic properties of ant "language" and estimate their cognitive skills.

        Ants are good candidates for studying general rules of natural communication and cognition, because these insects
        are known to combine highly integrative colony organization with sophisticated cognitive skills. Ants possess
        complex forms of communications, and they are known to be able to use a large variety of communication means for
        attracting their nestmates to a food source \cite{HW2}. It has remained unclear for a long time
        whether ants can use distant homing, that is, whether they can transfer messages about remote events without other
        cues such as scent trail or direct guiding. In this aspect, the so-called tactile (or antennal) "code" has been
        discussed since 1899, when it was   first hypothesised such an information transmission system in ants \cite{Wa}. However,
        the numerous attempts to decipher ants' "tactile language" have not given the desired results
        (for a review see \cite{Re3}). At the same time, it is clear that highly social ant species possess the
        necessary prerequisites for complex communication. Experimental studies revealed sophisticated forms of social
        learning in ants \cite{Re1}. However, methodological limitations have hampered the progress of studying
        "linguistic potential" of ants' communication, and the problem of distant homing in these insects has not been
        solved before our experiments.

          The experimental paradigm of our approach is simple. All we need to do is to establish a situation where ants
          must transfer a specific amount of information to each other. The crucial idea of the first scheme of
          experiments is that we know exactly the quantity of information to be transferred. To organize the process of
          information transmission between ants, a special maze has been used, called a "binary tree"
          \cite{RR1}, where the number and sequence of turns towards the goal corresponds to the amount
          of the information to be transferred.
          In another series of experiments ants had to transfer the information
          about the number of a branch in comb-like "counting mazes".

      It has been firstly demonstrated that group-retrieving Formica species possess distant homing, and they are able
      to pass meaningful messages. We also succeeded in studying important properties of ants' cognitive capacities,
      namely their ability to grasp regularities, to use them for coding and "compression" of information, and to add
      and subtract small numbers to optimize their messages.

      The obtained results demonstrate what Information Theory
      can furnish for explorers of communication and intelligence in social animals. This new experimental paradigm
      provides a way for studying important characteristics of animal communication which have not been accessible
      to study before, such as the rate of information transmission and the potential flexibility of communication systems.
      We also succeeded in studying some important properties of ants' intelligence, namely, their ability to grasp
      regularities and use them for optimization their messages.

\section{Using Shannon entropy and Kolmogorov Complexity to study communicative  system in ants: the binary tree experiments   }

The experiments based on Shannon entropy present a situation in
which, in order to obtain food, ants have to transmit certain
information which is quantitatively known to the researcher. This
information concerns the sequence of turns towards a trough with
syrup. The laboratory maze "binary tree" is used where each "leaf"
of the tree ends with an empty trough with the exception of one
filled with syrup. The leaf on which to place the filled trough
was chosen randomly by tossing a coin for each fork in the path.
The simplest design is a tree with two leaves, that is, a Y-shaped
maze. It represents one binary choice which corresponds to one bit
of information. In this situation a scouting animal should
transmit one bit of information to other individuals: to go to the
right (R) or to the left (L) - see Fig. 1. In other experiments
the number of forks of the binary tree increased to six. Hence,
the number of bits necessary to choose the correct way is equal to
the number of forks, that is, turns to be taken (Fig 2 shows a
labyrinth with 3 forks).

    The use of ideas of Shannon Entropy allowed the presence of potentially
    unlimited numbers of messages in ant "language" to be demonstrated and estimates
    of the rate of information transmission (approximately 1 bit/min.) to be made.
    We also succeeded in studying some properties of ants' intelligence, namely,
    their ability to memorize and to use simple regularities, thus compressing the
    information available. The latter experiments were based on the ideas of Kolmogorov complexity.

\subsection{ The experimental scheme }

 \begin{figure}   \label{Fig1} \centering \includegraphics[scale=0.5]{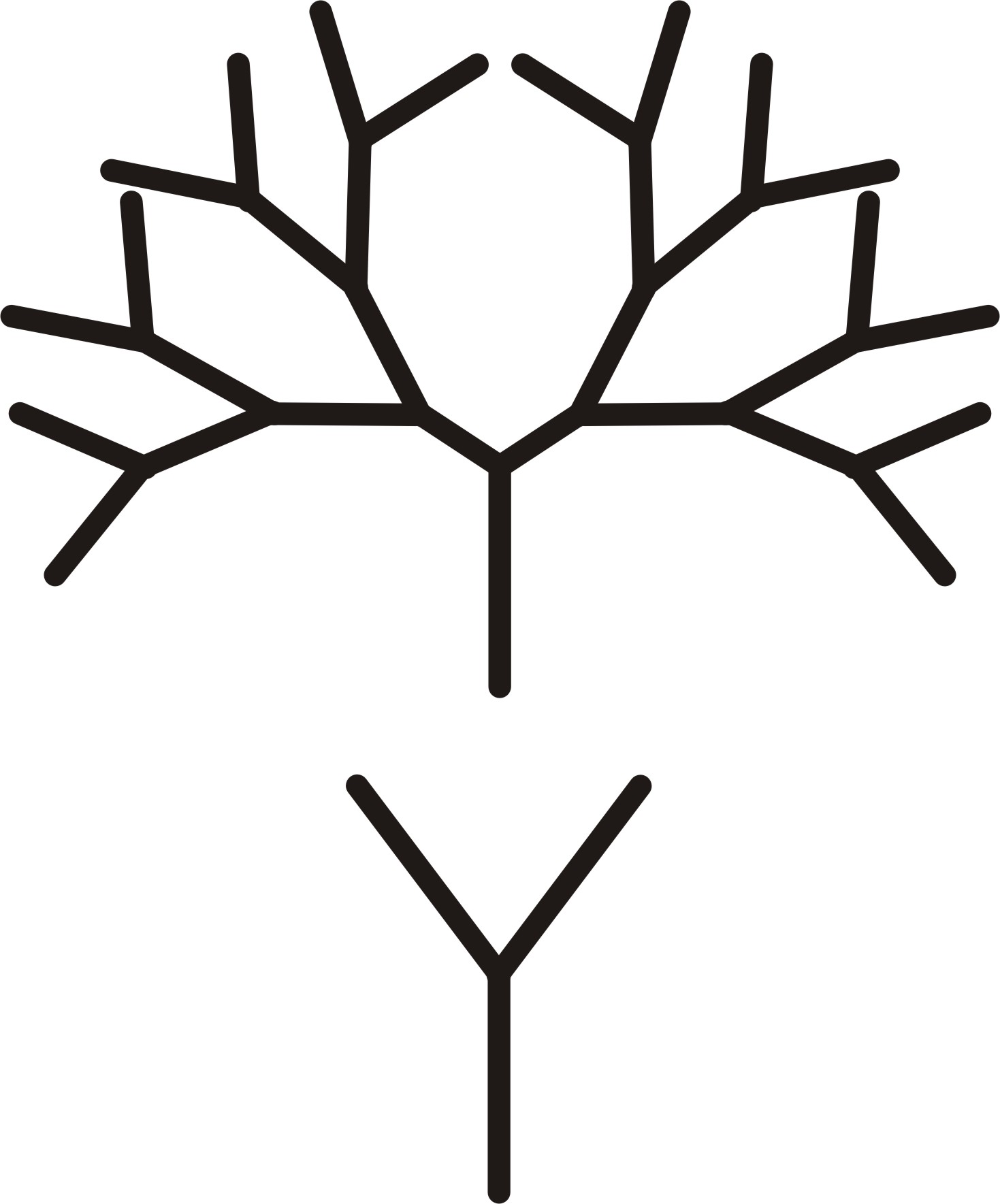}
        \caption{The maze "binary tree" with one fork and four forks.}\end{figure}

In the "binary tree" experiments ants were confronted with a
rather complex life-or-death task: they could obtain food only in
a "binary tree" maze and only once every 2 - 3 days. Ants
therefore were hungry and extremely motivated to obtain some food.
They had to search for the food placed on one of several "leaves"
of the "binary tree" maze (Figs. 1, 2). During different years
three colonies of \emph{Formica polyctena} and two of \emph{F.
sanguinea } were used. Ants lived in the $2 \times 1.5$-meters
laboratory arena, in a transparent nest that made it possible for
their activity to be observed. The arena was divided into two
sections: a smaller one containing the nest, and a bigger one with
an experimental system (Fig. 2). The two sections were connected
by a plastic bridge that was removed from time to time to modify
the set-up or to isolate the ants. To prevent access to the food
in the maze by a straight path, the set-up was placed in a bath of
water, and the ants reached the initial point of the binary tree
by going over a second small bridge.

    The laboratory colonies consisted of about 2000 individuals each. All actively
    foraging ants were individually marked with colored paint. The laboratory colonies
    were found to include teams of constant membership which consisted of one scout
    and three to eight recruits (foragers): the scout mobilized only members of its
    team to the food. The composition of the teams was revealed during special run-up
    experiments consisting of familiarization trials lasting as long as two or three weeks
    (see details in \cite{Re3}). In total, 335 scouts along with their teams were used
    in all experiments with the binary tree. In each trial one of the scouts that were actively
    moving on the experimental arena at that moment was placed on a leaf of the binary tree that
    contained a trough with the food, and then it returned to the nest by itself. All contacts
    between the scout and its team were observed and time duration was recorded each time.

    \begin{figure}   \label{f2} \centering \includegraphics[scale=0.15]{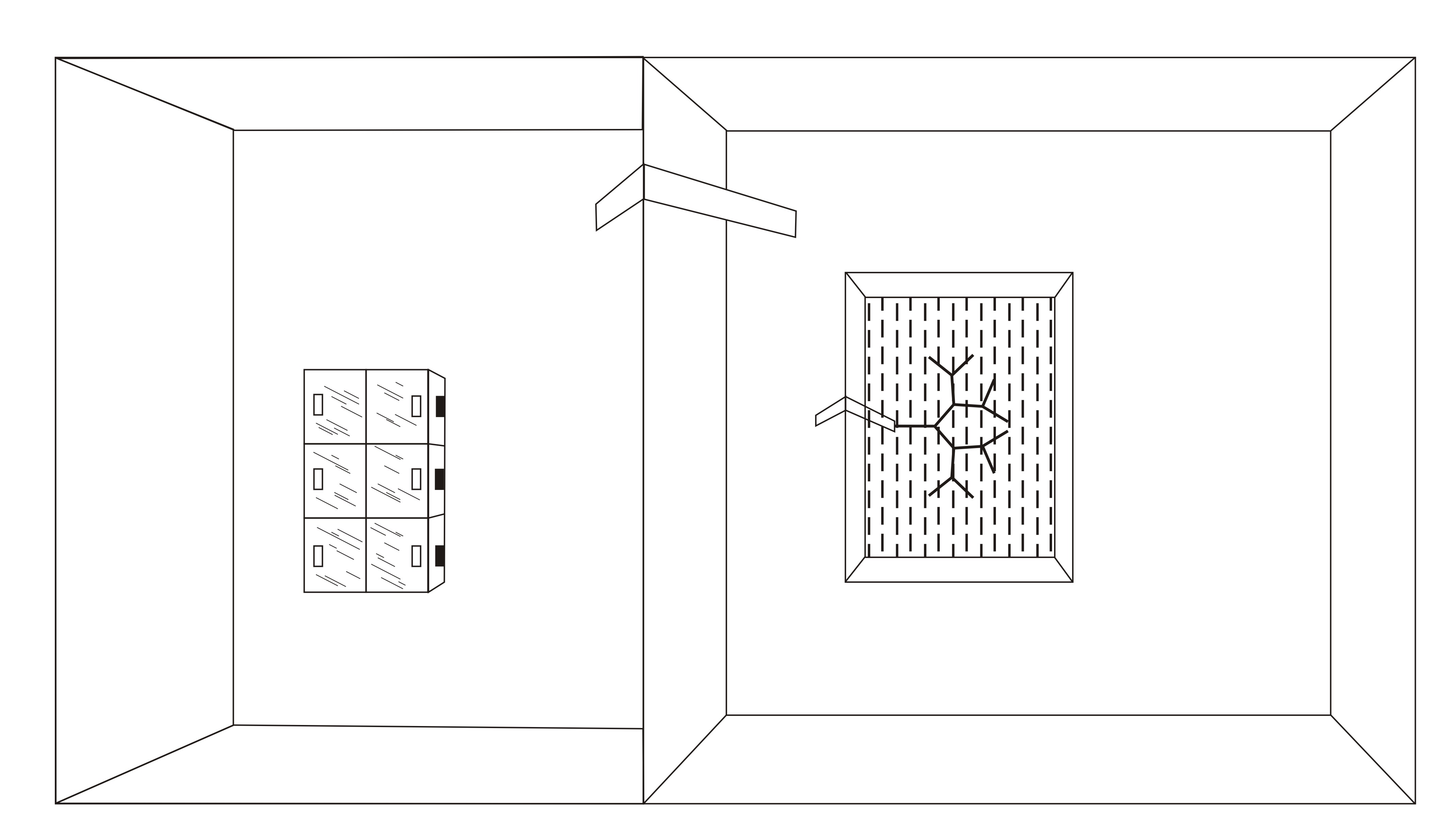}
     \caption{The laboratory arena with the maze "binary tree".}\end{figure}

     All experiments were so devised as to eliminate all possible cues that could help the ants
     to find the food, except information contact with the scout. To avoid the use of an odor track,
     the experimental set-up was replaced by an identical one when the scout was in the nest or on
     the arena contacting its group. All troughs in the fresh maze contained only water to avoid
     the possible influence of the smell of syrup. If the group reached the correct leaf of the
     binary tree, they were immediately presented with the food.
     The scout had to make up to four trips before it was able to mobilize its group of foragers.
     Usually members of the team had already left the nest after the scout's first trip and were waiting
     on the arena for its return. Returning to the group, the scout contacted one to four foragers in turn,
     sometimes two of them simultaneously. Contacts were followed by numerous antennal movements.
     The experiments were designed to investigate the characteristics of distant homing, so after the scout
     had contacted its team, it was isolated for a while, and the foragers had to search for the food by themselves.
     This process will be described in more details in the next section. Here it is important to note that the
     composition of the working teams remained constant in each colony from several days to several weeks, that is,
     during periods when a given scout was actively working (see detailed tables in \cite{RR1}).
     It is notable that in both \emph{F. polyctena} and \emph{F. sanguinea}, not all of the scouts managed to memorize  the way to
     the correct leaf of the maze even after they had passed their "final exams" during the run-up trials. The number
     of scouts that succeeded in memo-rising the way decreased with increasing complexity of the task. In the case
     of two forks all active scouts and their groups (up to 15 per colony) were successful whereas in the case
     of six forks, only one or two coped with the task.

      During the experiments each scout was placed on the trough containing food, and after the scout had eaten
      it returned to the nest on its own. In all cases of mobilization of the group the duration of the contact
      between the scout and the foragers was measured in seconds. The contact was considered to begin when the scout
      touched the first forager ant, and to end when the first two foragers left the nest for the maze. When the scout
      repeatedly returned to the trough alone, each of its contacts with foragers was measured. Only the duration
      of the contact that was followed by the foragers' leaving the nest was taken into account. These contacts were
      hypothesized to be "informative", and they differed sharply in duration from other contacts: all were more than
      30 seconds. As a rule, all of the previous contacts between scouts and foragers were brief (about 5 seconds)
      and were made for the exchange of food.
     During each series of experiments with the trough placed on the $i-$th leaf of the binary tree, all teams that
     were active on that day worked successively. While the trial was going on, the bridge leading to the working
     part of the arena was taken away, so as not to let members of other teams to go there. While the scout was
     inside the nest, the whole maze was replaced by a fresh one with all troughs empty. Foragers were presented
     with the syrup if they reached the correct leaf of the binary tree.

\subsection{ Information transmission by distant homing in ants:
statistical proof}

 Before analyzing ants' "linguistic potential" and their ability to use rules for compression of information
 we consider the evidence of information transmission from the scouts to the foragers, which came from two
 sets of data: first, from statistical analysis of the number of faultless findings of the goal by a group,
 and second, from a special series of control experiments with "uninformed" ("naive") and "informed" foragers.
       The statistical analysis of the number of faultless findings of the goal was carried out by comparing
       the hypothesis $H_0$ (ants find the leaf containing the food by chance) with the hypothesis $H_1$ (they find
       the goal thanks to the information obtained), proceeding from the fact that the probability of finding
       the correct way by chance when the number of forks is $i$ is $(1 / 2)^i$. We analyzed different series of
       experiments (338 trials in sum), separately for 2, 3, 4, 5, and 6 forks. In all cases $H_0$ was rejected
       in favor of $H_1, P < 0.001$ (see \cite{RR2}), thus unambiguously demonstrating information
       transmission from scouts to foragers.

     The control experiments were organized so as to compare searching results of the ants that had and had
     not previous possibility to contact the scout (the "informed" and "naive" ants, respectively). The
     "naive" and "informed" ants were tested one by one. Each ant was allowed to search for the food for 30 min.
     In Table 1 the time spent on searching the trough by "informed" and "uninformed" \emph{Formica pratensis} are compared
     \cite{RR4,No}. For every trial, Wilcoxon's non-parametric test was used
     \cite{HW1} to test the hypothesis $H_0$ (data from both samples follow the same distribution)
     against $H_1$ (they follow different distributions) at significance level $0.01$.
     We concluded that the duration of searching time is essentially smaller in those ants that had previously contacted the scout.
      These data demonstrate that scouts transfer information about the discovered food to foragers by means of distant homing.
    \begin{table}[h]\label{tab1}\caption{Comparison of duration of searching the trough by "uninformed" (U)
     \emph{F. pratensis} ants and individuals that previously contacted with the successful scout ("Informed",I)}
\centering
 \begin{tabular}{|c|c|c|c|c|}
\hline
 Sequence of the turns    & Ants (U/I)  & Mean (second)& Amounts of sampling & P  \\
   \hline
 RRRR & U & 345.7 & 9& $< 0.01 $\\
  & I & 36.3 &9 &  \\ \hline
  LLLL & U & 508.0 & 9& $< 0.01 $\\
  & I & 37.3 &9 &  \\ \hline
  LRRL & U & 118.7 & 7& $< 0.01 $\\
  & I & 16.6 &7 &  \\ \hline
  RRRR & U & 565.9 & 7& $< 0.01 $\\
  & I & 16.3 &7 &  \\ \hline
\end{tabular}
\end{table}

\subsection{ The rate of information transmission in ants}

We now can evaluate the rate of information transmission in ants.
To do this, observe that the quantity of information (in bits)
necessary to choose the correct route in the maze equals $i$, the
depth of the tree (the number of turns to be taken), that is, $
\log_2 n$ where $n$ is the number of leaves. One can assume that
the duration of the contacts between the scouts and foragers ($t$)
is $a  i + b$, where $i$ is the number of turns (the depth of the
tree), $a$ is the rate of information transmission (bits per
minute), and $b$ is an introduced constant, since ants can
transmit information not related directly to the task, for
example, the simple signal "food". Besides, it is not ruled out
that a scout ant transmits, in some way, the information on its
route to the nest, using acoustic or some other means of
communication. In this context, it is important that the route
from the maze to the nest on the arena was in all experiments
approximately the same. Being highly motivated, scouts hurried on
to the nest in a beeline, and, therefore, the time before they
made antennal contact with the foragers in the nest, which the
scout could hypothetically use for message transmission, was
approximately the same and did not depend on the number of turns
to be taken in the maze.

     From the data obtained, the parameters of linear regression and the sample correlation coefficient
     ($r$) can be evaluated. The rate of information transmission (a) derived from the equation
     $t = a  i + b$ was 0.738 bits per minute for \emph{F. sanguinea} and 1.094 bits per minute for \emph{F. polyctena}.
     The rate of information transmission is relatively small in ants.

     To estimate the potential productivity of ants' "language",
     let us count the total number of different possible routes to the trough.
     In the simplest binary tree with one fork there are two leaves and therefore two
     different routes. In a tree with two forks there are $2^2$ routes, with three forks $2^3$ routes,
     and with six forks, $2^6$ routes; hence, the total number of different routes is equal to
     $2 + 2^2 + 2^3 + \ldots + 2^6 =
126$. This is the number of messages the ants must be able to pass
in order to pass the information about the food placed on any leaf
of the binary tree with 6 forks.

           \subsection{ The Kolmogorov complexity and data compression in the ants language }

     Another series of experiments with the binary tree was inspired by the concept of Kolmogorov
     complexity and was designed to check whether highly social ant species possess such an important
     property of intelligent communications as the ability to grasp regularities and to use them for encoding
     and "compressing" information. This concept is applied to words (or text) composed of the letters of any alphabet,
     for example, of an alphabet consisting of two letters: L and R. We interpret a word as a sequence of
     left (L) and right (R) turns in a maze. Informally the complexity of a word (and its uncertainty)
     equates to its most concise description, according to Kolmogorov \cite{K}.
     For example, the word "LLLLLLLL" can be represented as "8 L", the word "LRLRLRLR" as "4LR",
     while the "random" word of shorter length "LRRLRL" probably cannot be expressed more concisely,
     and this is the most complex of the three.

    We analyzed the question of whether ants can use simple regularities of a "word" to compress it.
    It is known that Kolmogorov complexity is not algorithmically computable. Therefore, strictly speaking,
    we can only check whether ants have a "notion" of simple and complex sequences. In our binary tree maze,
    in human perception, different routes have different complexities.

    In one particular series of experiments,
    \emph{F. sanguinea} ants were presented with different sequences of turns.
    Testing the hypothesis $H_0$ (the time for transmission of information does not depend on
    the text complexity) against the hypothesis $H_1$ (that time actually depends on it) allowed us
    to reject $H_0$, thus showing that the more time ants spent on the information transmission,
    the more information - in the sense of Kolmogorov complexity - was contained in the message \cite{RR96}.

    Let us test these hypotheses formally.
    There are seven sequences of turns of equal length (lines 5-8
    and 13-15 at the table 2). The total number of turn sequences
    orders, according to the duration of the transmission is $7!$
    of which $2! 2! 3! $   are in the line with $H_0$. The
    probability of obtaining such an order according to $H_0$ is
    very small: $ (2! 2! 3!) / 7! $  $= 1/210$. Thus, we accept
    the hypothesis  $H_1$: the simpler the text the less time for
    information transmission.

    It is interesting that the ants began to use regularities to compress only quite large "words".
    Thus, they spent from 120 to 220 seconds to transmit information about random turn patterns on
    the maze with five and six forks and from 78 to 135 seconds when turn patterns were regular.
    On the other hand, there was no essential difference when the length of sequences was less than 4 (Table 2).

     \begin{table}[ht]\label{tab1}
\centering
     \caption{Duration of transmitting information on the way to the trough
by \emph{F.sanguinea} scouts to foragers (no.1-8 regular turn
pattern;
 no. 9-15 random turn pattern)
}
 \begin{tabular}{|c|c|c|c|}
\hline
 No   & Sequence  & Mean  duration (sec)&  Numbers of  experiments \\
   \hline
 1 & LL & 72 & 18\\
   \hline
   2 & RRR & 75 &  15\\
   \hline
   3 & LLLL & 84 & 9\\
   \hline
   4 & RRRRR & 78 & 10\\
   \hline
   5 & LLLLLL & 90 & 8\\
   \hline
   6 & RRRRRR & 88& 5\\
   \hline
   7 & LRLRLR & 130 & 4\\
   \hline
   8 & RLRLRL& 135 & 8\\
   \hline
   9& LLR & 69 & 12\\  \hline
   10 & LRLL & 100 & 10\\
   \hline
   11& RLLR & 120& 6\\
   \hline
   12 & RRLRL & 150 & 8\\
   \hline
   13 & RLRRRL & 180 & 6\\
   \hline
   14 & RRLRRR & 220 & 7\\
   \hline
   15 & LRLLRL & 200 & 5\\
   \hline

  \end{tabular}
\end{table}

    These results enable us to suggest that ants not only produce a large number of messages but can
    use rule extraction in order to optimise their messages. The ability to grasp regularities and to
    use them for coding and "compression" of information can be considered as one of the most important
    properties of language and its carriers' intellect. Thus we can conclude that the ants' communication
    system is rational and flexible.

\section{ The ants numerical competence}

Basic number-related skills, that is, knowledge of quantities and
their relations, is, perhaps, one of the highest properties of
cognition. Recent studies have demonstrated some species as being
able to judge about numbers of stimuli, including things, and
sounds, and maybe smells. For example, lions can count roaring
that comes from individuals who are not members of the pride
\cite{MPP}; honey bees are able to use the number of landmarks as
one of the criteria in searching for food sources \cite{CG}. There
are many other examples that come from different animal species,
from mealy beetles \cite{SFFD}  to elephants \cite{IKSH}; however,
we are still lacking an adequate "language" for comparative
analysis. The main difficulty in comparing numerical abilities in
humans and other species is that our numerical competence is
closely connected with abilities for language usage and for
symbolic representation.

    We elaborated an experimental paradigm for studying ants' numerical competence
\cite{RR3}. A scouting ant has to transfer the information about a
number (in our case, the index number of a branch of a maze) to
its nest mates. Quantitative characteristics of the ants'
communications were used for investigating their ability to count.
The main idea of this experimental paradigm is that experimenters
can judge how ants represent numbers by estimating how much time
individual ants spend on "pronouncing" numbers, that is, on
transferring information about index numbers of branches.

\subsection{ The comb-like set - ups }

The experiments were based on a procedure similar to the binary
tree study. Ant scouts were required to transfer to foragers in a
laboratory nest the information about which branch of a special
"counting maze" they had to go to in order to obtain syrup.
"Counting maze" is a collective name for several variants of
set-ups. All of them serve to examine how ants transfer
information about index numbers of branches by means of distant
homing.

\begin{figure}   \label{f3} \centering \includegraphics[scale=0.6]{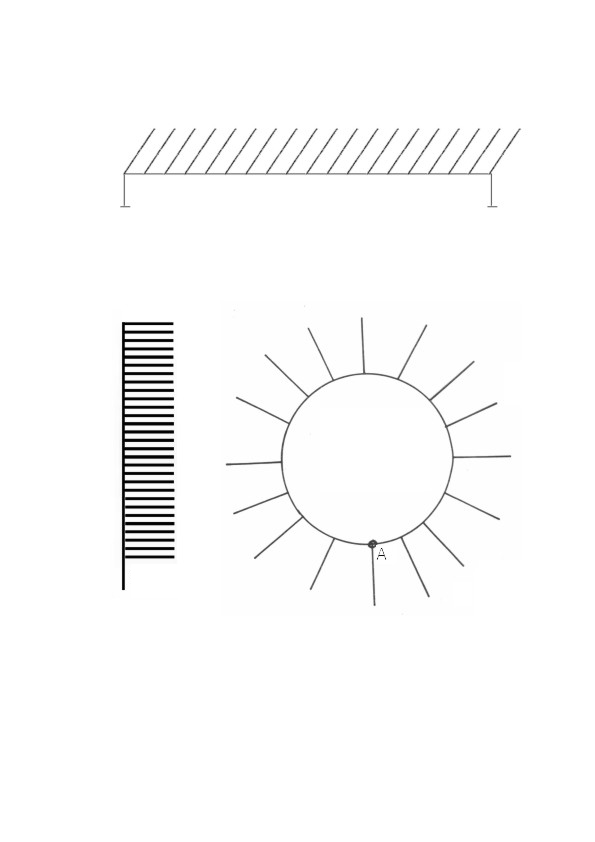}
 \caption{The comb-like set-ups for studying numerical competence in ants: a horizontal trunk, a vertical trunk and a circle.
}\end{figure}

The first variant of the counting maze is a comb-like set-up
consisting of a long horizontal plastic trunk with 25 to 60
equally spaced plain plastic branches, each of them 6 cm in length
(Fig.3). Each branch ended with an empty trough, except for one
filled with syrup. Ants came to the initial point of the trunk
over a small bridge. The second variant is a set-up with
vertically aligned branches. In order to test whether the time of
transmission of information about the number of the branch depends
on its length as well as on the distance between the branches, one
set of experiments was carried out on a similar vertical trunk in
which the distance between the branches was twice as large, and
the branches themselves were three times and five times longer
(for different series of trials). The third variant was a circular
trunk with 25 cm long branches.


    Similarly to the binary tree study, ants were housed in a laboratory arena
    divided into two parts, one containing a plastic nest with a laboratory ant
    colony and another containing one of the variants of the counting maze.
    During different years two laboratory colonies of \emph{F. polyctena} were used in
    this set of experiments. Each series of experiments was preceded by the run-up
    stage consisting of familiarization trials. In order to force a scout to transfer
    the information about food to its nest mates we showed it the trough containing syrup
    (placing the scout directly on the trough) and then let it return to the nest.
    After allowing it to contact the foragers within the nest, the scout was removed and
    isolated for a while, so that the foragers had to search for the food by themselves,
    without their guide.

     Again, similar to the binary tree study, the experiments with counting mazes were
     devised so as to eliminate all possible ways for the members of each foraging team
     to find a goal, except by distant homing, i.e., an information contact with their scout.
     The set-up was replaced with a fresh one, with all troughs filled with water,
     while the scout was in the nest; if the foraging team reached the correct branch
     in a body, then the water-filled trough was replaced with one with syrup; thus foragers
     had to rely solely on the information from the scout.

\subsection{ The ants' ability to transfer the information about numbers of objects}

The findings concerning number-related skills in ants are based on
comparisons of duration of information contacts between scouts and
foragers which preceded successful trips by the foraging teams.
Duration of contacts of the scout with its team was measured when
the scout returned from the experimental set-up, loaded with both
syrup and information. In total, 32 scout - foragers teams worked
in three kinds of set-ups. The teams left the nest after they were
contacted by scouts and moved towards the trough by themselves 152
times (recall that the scouts were removed and the set-ups were
replaced). In 117 cases the team immediately found the correct
path to the trough, without making any wrong trips to empty
troughs. In the remaining cases, ants came to the empty troughs,
and began looking for food by checking neighboring branches.

     Since all set-ups had no fewer than 25 branches, the probability of finding
     the correct trough by chance is not more than $1 / 25$. Thus, the success ratio
     which was obtained experimentally can only be explained by information transmission
     from the scouts. The probability of finding the food-containing trough by chance in
     117 cases out of 152 is less than $10^{-10}$. In addition, in control experiments ants, including
     scouts placed in the set-up, without information on which trough contained food
     usually failed to find the food, even though they actively searched for it.

        \begin{table}[ht]\label{tab1}
     \caption{The results of experiments in the "vertical trunk 1" with \emph{F. polyctena.}
}
\centering
 \begin{tabular}{|c|c|c|c|}
\hline
 No   & Number of food- & Duration of scout &the scout's  "name"   \\
    &   containing branch & -forager contact (sec)   & \\
   \hline
 1 & 10 & 42 & I\\
   \hline
   2 & 10 & 40 & II\\
   \hline
   3 & 10 & 45 &  III\\
   \hline
   4 & 40 & 300 & II\\
   \hline
   5 & 40 & 280 &IX\\
   \hline
   6 & 13  &90 & II\\
   \hline
   7 & 13  &98 & I\\
   \hline
   8 & 28  &110& III\\
   \hline
   9& 28 & 120& X\\  \hline
   10 & 20 & 120&  X\\
   \hline
   11& 20 & 110 & III\\
   \hline
   12 & 35  &260 & III\\
   \hline
   13 & 35 & 250 & X\\
   \hline
   14 & 30 & 160& I\\
   \hline
   15 & 30& 170 & III\\
   \hline

  \end{tabular}
\end{table}

     Data obtained on the vertical trunk are shown in Table 3 as an example. It turned out
     that the relation between the index number of the branch and the duration of the contact
     between the scout and the foragers is well described by the equation $t = a  i + b$
     for different set-ups which are characterized by different shapes, distances between
     the branches and lengths of the branches. The values of parameters $a$ and $b$ are close
     and do not depend either on the lengths of the branches or on other parameters.
     The correlation coefficient between $t$ and $i$ was high for different kinds of counting mazes; see  Table 4.

     \begin{table}[ht]\label{tab4}
\centering
\caption{Values of correlation coefficient ($r$) and regression
($a, b$) coefficients for vertical trunk (vert), horizontal trunk
(horiz), and circle in the experiments with\emph{ F. polyctena} }
 \begin{tabular}{|c|c|c|c|c|c|}
\hline
 Type of setup  & Sample size& Numbers of   branches
& r & a  & b  \\
 \hline
 Vert.1 & 15 & 40& 0.93 & 7.3  & -28.9\\
\hline
Vert.2 & 16 & 60& 0.99 & 5.88  & -17.11\\
\hline
Horiz.1 & 30& 25& 0.91 & 8.54  & -22.2\\
\hline
Horiz.2 & 21& 25& 0.88 & 4.92  & -18.94\\
\hline
Circle & 38& 25& 0.98 & 8.62  & -24.4\\
\hline
\end{tabular}
\end{table}

     All this enables us to suggest that the ants transmit information solely concerning the
     index number of the branch.

      It is interesting that quantitative characteristics of the ants' "number system" seem to be close,
      at least outwardly, to some archaic human languages: the length of the code of a given number
      is proportional to its value. For example, the word "finger" corresponds to 1, "finger, finger"
      to the number 2, "finger, finger, finger" to the number 3 and so on. In modern human languages
      the length of the code word of a number i is approximately proportional to $ \log i$ (for large $i$'s),
      and the modern numeration system is the result of a long and complicated development.

\subsection{ The ants' ability to add and subtract small numbers}

   There are some experimental evidence that two years old human children, rhesus monkeys and chimpanzees can operate
   with addition and subtraction of up to 4. The experiments are based on games where subjects should
   demonstrate their abilities to judge about the number of items after the addition or removal \cite{GaGe}.

   We elaborated a new experimental paradigm of studying ants' "arithmetic" skills based on a fundamental idea of information
   theory, which is that in a "reasonable" communication system the frequency of usage of a message and its length must
   correlate. The informal pattern is quite simple: the more frequently a message is used in a language, the shorter
   is the word or the phrase coding it. This phenomenon is manifested in all known human languages.

     The main experimental procedure was similar to other experiments with counting mazes. In various years four
     colonies of \emph{F. polyctena} were used in this set of experiments. The scheme of the experiments was as follows.
     Ants were offered a horizontal trunk with 30 branches. The experiments were divided into three stages, and at
     each of them the regularity of placing the trough with syrup on branches with different numbers was changed.
     At the first stage, the branch containing the trough with syrup was selected randomly, with equal probabilities
     for all branches. So the probability of the trough with syrup being placed on a particular branch was $1 / 30$.
      At the second stage we chose two "special" branches A and B (N 7 and N 14; N 10 and N 20; and N 10 and N 19
      in different years) on which the trough with syrup occurred during the experiments much more frequently
      than on the rest - with a probability of $1 / 3$ for "A" and "B", and $1 / 84$ for each of the other 28 branches.
      In this way, two "messages" - "The trough is on branch A" and "The trough is on branch B" - had a much higher
      probability than the remaining 28 messages. In one series of trials we used only one "special" point A (the branch N 15).
      On this branch the food appeared with the probability of $1 / 2$, and $1 / 58$ for each of the other 29 branches.
      At the third stage of the experiment, the number of the branch with the trough was chosen at random again.

     \begin{figure}   \label{ris2} \centering \includegraphics[scale=1]{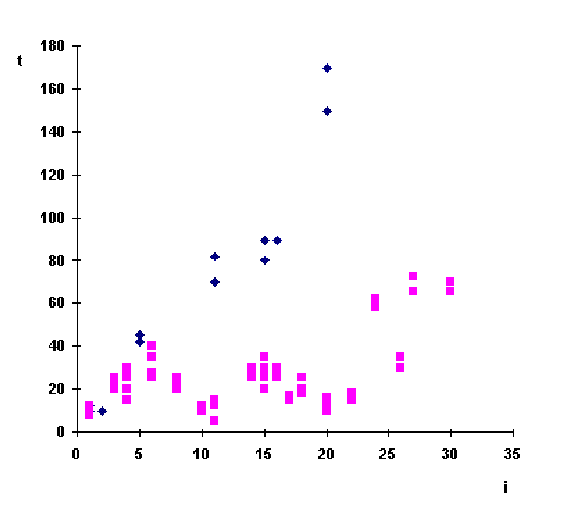}
     \caption{Dependence of the time ($t$; measured in seconds)
of transmission of information about the number of the branch
having food on its ordinal number ($i$) in the first and the third
series of experiments in the ant \emph{Formica polyctena}.
Diamonds, the time taken for transmission of information at the
first stage; Squares, the same it the third stage.}\end{figure}

    Now let us consider the relationship between the time which the ants spent to transmit the information about
     the branch containing food, and its number.
     The information obtained at the first and third stages of the
     experiments are shown on the graph (Fig. 4) in which the time of the scout's contact with foragers $(t)$ is plotted
     against the number $(i)$ of the branch with the trough. At the first stage the dependence is close to linear.
     At the third stage, the picture was different: first, the information transmission time was very much reduced,
     and, second, the dependence of the information transmission time on the branch number is obviously non-linear:
     depression can be seen in the vicinities of the "special" points (10 and 20). So the data demonstrate that the
     patterns of dependence of the information transmission time on the number of the food-containing branch at the
     first and third stages of experiments are considerably different. Moreover, in the vicinities of the "special"
     branches, the time taken for transmission of the information about the number of the branch with the trough is,
     on the average, shorter. For example, in the first series, at the first stage of the experiments the ants took
     70 - 82 seconds to transmit the information about the fact that the trough with syrup was on branch N 11, and
     8 - 12 seconds to transmit the information about branch N 1. At the third stage it took 5 - 15 seconds to transmit
     the information about branch N 11. These data enable us to suggest that the ants have changed the mode of presenting
     the data about the number of the branch containing food.

 \begin{table}[ht]\label{tab1}
\centering
     \caption{Dependence of the time of information transmission ($t$) on
the distance from the branch with a trough to the nearest
"special" branch (special branches are 10 and 20) }
 \begin{tabular}{|c|c|c|}
\hline
 The number of  the branch having  & Distance to  the nearest & Times of  transmission of  \\
 food (experiments in different days,&special branch &information  about the  branch   \\
 consequently) & & number for  different scouts \\
& &  (sec) \\
  \hline
 26 & 6 & 35,30 \\
\hline 30 & 10 & 70,65 \\ \hline
27 &7 &65,72 \\
 \hline  24&4 &58,60,62 \\
 \hline 8 & 2&22,20,25 \\
 \hline  16 & 4&25,8,25 \\
 \hline 16  &4 &25 \\
 \hline 22 & 2& 15,18\\
 \hline18  &2 &20,25,18,20 \\
 \hline  15   & 5&30,28,35,30 \\
 \hline 20  &0 &10,12,10 \\
 \hline  6& 4&25,28 \\
 \hline 16 & 4& 30,25\\
 \hline 15 & 5& 20,25,20\\
 \hline  14& 4&25,28,30,26 \\
 \hline  17& 3&17,15 \\
 \hline  11&1 &10,12 \\
 \hline

\end{tabular}
\end{table}

     What about ants' ability to add and subtract small numbers? Analysis of the time duration of information transmission
     by the ants raises the possibility that at the third stage of the experiment the scouts' messages consisted of two parts:
     the information about which of the "special" branches was the nearest to the branch with the trough, and the information
     about how many branches away is the branch with the trough from a certain "special" branch. In other words, the ants,
     presumably, passed the "name" of the "special" branch nearest to the branch with the trough, and then the number
     which had to be added or subtracted in order to find the branch with the trough. That ant teams went directly to
     the "correct" branch enables us to suggest that they performed correctly whatever "mental" operation (subtraction
     or addition) was to be made.

    In order to verify this statistically, the coefficient of correlation was calculated between the time required for
    transmission of information about the trough being on the branch $i$ and the distance from $i$ to the nearest "special"
    branch. The results confirmed the hypothesis that the time for transmission of a message about the number of the
    branch is shorter when this branch is closer to any of the "special" ones. For this purpose, the data obtained at
    the third stage of the experiment were transformed to present them in the form shown in Table 5 where data of one
    year are given as an example. In this table we do not include branches that are close to the starting point of
    the set-up (N 1 - 4) because there is no need to use "arithmetic" for ants where rewarded branches are very close
    to the first one (in fact ants spent roughly the same time transmitting information about these branches: from 10 to 20 seconds).

           \begin{table}[ht]\label{tab1}
\caption{Values of correlation coefficient (r) in the experiments
with different "special" branches }
\begin{tabular}{|c|c|c|c|}
\hline
 Sample size  &  Numbers of  "special"  & $r$ for the  first stage & $r$ for the  third   stage
 \\
& branches&of the  experiments& of the experiments\\
 \hline
150&10,20 &0.95 &0.80 \\
 \hline
 92&10,19 &0.96 &0.91 \\
 \hline
 99&15 &0.99 &0.82 \\
 \hline

\end{tabular}
\end{table}

      It can be seen from Table 6 that the coefficients of correlation between the transmission time and the distance
      to the nearest special point have quite high values and they differ significantly from zero (at the confidence
      level of 0.99). So the results support the hypothesis that the time for transmission of a message about the
      number of the branch is shorter when this branch is close to either of the special ones. This, in turn, shows
      that at the third stage of the experiment the ants used simple additions and subtractions, achieving economy
      in a manner reminiscent of the Roman numeral system when the numbers 10 and 20, 10 and 19 in different series
      of the experiments, played a role similar to that of the Roman numbers V and X.

      Our interpretation is that ants of highly social group-retrieving species are able to add and subtract small
      numbers. This also indicates that these insects have a communication system with a great degree of flexibility.
      Until the frequencies with which the food was placed on different branches started exhibiting regularities,
      the ants were "encoding" each number ($i$) of a branch with a message of length proportional to $i$, which suggests unitary
      coding. Subsequent changes of code in response to special regularities in the frequencies are in line with a basic
      information-theoretic principle that in an efficient communication system the frequency of use of a message and
      the length of that message are related.

\section*{Acknowledgements} Research  was supported  by Russian Foundation for Basic Research (grants
     09-07-00005 and 08-04-00489)

\bibliographystyle{mdpi}

\begin{thebibliography}{1}

\bibitem{BH}
Boysen, S. T.; Hallberg, K. I. Primate numerical competence:
contributions toward understanding nonhuman cognition. Cognitive
Science 2000,24, 3, 423-443.

\bibitem{BKK}
Bugnyar, T.; Kijne, M.; Kotrschal, K. Food calling in
ravens: are yells referential signals?. Animal Behaviour 2001, 61,
949-958.

\bibitem{SFFD}
Carazo, P.; Font, E.; Forteza-Behrendt, E.; Desfilis, E.
Quantity discrimination in Tenebrio molitor: evidence of
numerosity discrimination in an invertebrate? Animal Cognition 2009,
12, 462-470.

\bibitem{CG}
Chittka, L.; Geiger, K. Can honeybees count landmarks?
Animal Behaviour 1995, 49, 159-164.

\bibitem{EB}
Evans, W. E.; Bastian, J. Marine mammal communication;
social and ecological factors. In: The Biology of Marine Mammals,
ed. H. T. Andersen. New York: Academic Press,1969,425-475.

\bibitem{Fr1}
Frisch, K. von. Uber die Sprache der Bienen. Zoologische
Jahrbucher - Abteilung fur Allgemeine Zoologie und Physiologie der
Tiere 1923, 40: 1-119.

\bibitem{Fr2}
Frisch, K. von. The dance language and orientation of bees.
- Harvard University Press, Cambridge, MA, 1967, 566 pp.

\bibitem{GaGe}
Gallistel, C. R.; Gelman, R. Preverbal and verbal
counting and computation. Cognition 1992, 44, 43-74

\bibitem{GG}
Gardner, R. A., Gardner, B. T. Teaching sign language to a
chimpanzee. Science 1969, 165, 664-672.

\bibitem{HAEH}
Herman, L. M.; Abichandani, S. L.; Elhajj, A. N.; Herman, E. Y.
K.; Sanchez, J. L.; Pack, A. A. Dolphins (Tursiops
truncatus) comprehend the referential character of the human
pointing gesture.Journal of Comparative Psychology 1999, 113, 1-18.

\bibitem{HW1}
Hollander, M.; Wolf, D.A. Nonparametric statistical methods.
- John Wiley and Sons, New York, 1973.

\bibitem{HW2}
Holldobler, B.; Wilson, E. O. The ants. Cambridge: The
Belknap Press of Harward University Press, 1990.

\bibitem{HM}
Hollen, L.I.; Manser, M.B. Ontogeny of alarm call responses
in meerkats, Suricata suricatta: the roles of age, sex and nearby
conspecifics. Animal Behaviour 2006,72,1345-1353.

\bibitem{IKSH}
Irie-Sugimoto, N.; Kobayashi, T.; Sato, T.; Hasegawa, T.
Relative quantity judgment by Asian elephants (Elephas maximus).
Animal Cognition 2009,12,193 -199.

\bibitem{K}
Kolmogorov, A. N. Three approaches to the quantitative
definition of Information. Problems in Information Transmission,
1965,1,1-7.

\bibitem{MPP}
McComb, K.; Packer, C.; Pusey, A. Roaring and numerical
assessment in contests between groups of female lions, Panthera
leo. Animal Behaviour, 1994, 47, 379- 387.

\bibitem{MH}
McCowan, B.; Hanser, S. F.; Doyle, L. R. Quantitative Tools for Comparing Animal
Communication Systems: Information Theory Applied to Bottlenose Dolphin Whistle
Repertoires. Animal Behaviour 1999, 57, 409-419.

\bibitem{Me}
Menzel, E. W. Communication about the envirinment in a group
of young chimpanzees. Folia Primatologica 1971,15, 220-232.


\bibitem{M}
Michelsen, A. The transfer of information in the dance
language of honeybees: progress and problems.Journal of
Comparative Physiolology, A, Sensory, Neural and Behavioural
Physiology 1993,173, 135-141.

\bibitem{No}
Novgorodova,T.A. Experimental investigation of foraging
modes in Formica pratensis  (Hymenoptera, Formicidae) using
"Binary Tree" maze. Entomological Review 2006, 86, 287-293.

\bibitem{Pe}
Pepperberg, I. M. The Alex studies. Cambridge, MA: Harvard
University Press, 1999.

\bibitem{Re1}
Reznikova, Zh. Interspecific communication among ants.
Behaviour 1982, 80, 84-95.

\bibitem{Re1A}
Reznikova, Zh. Dialog with black box: using Information
Theory to study animal language behaviour. Acta ethologica 2007-a, 10,
1-12.

\bibitem{Re2}
Reznikova, Zh. Animal Intelligence: From individual to
social cognition. Cambridge University Press, Cambridge MA, 2007-b.

\bibitem{Re3}
Reznikova, Zh. 2008. Experimental paradigms for studying cognition
and communication in ants (Hymenoptera: Formicidae). Myrmecological News 2008,
11, 201-214.

\bibitem{rr0}
Reznikova, Zh., Ryabko, B. Investigations of ant language by
methods of Information Theory. Problems of Information Theory
1986, 21, 3 , 103-108.

\bibitem{RR1}
Reznikova, Zh.; Ryabko, B. Experimental study of the ants
communication system with the applicatiob of the Information
Theory approach. Memorabilia Zoologica 1994, 48, 219-236.

\bibitem{RR2}
Reznikova, Zh.; Ryabko, B. Transmission of information
regarding the quantitative characteristics of objects in ants.
Neuroscience and Behavioural Psychology 1996, 36, 396-405

\bibitem{RR3}
Reznikova, Zh.; Ryabko, B. Experimental study of ant capability
for addition and subtraction of small numbers. Neuroscience and
Behavioral Physiology 1999, 49, 2-21.

\bibitem{RR4}
Reznikova, Zh.; Ryabko, B. In the shadow of the binary tree: Of
ants and bits. In: Proceedings of the 2nd Internat. Workshop of
the Mathematics and Algorithms of Social Insects, ed. C. Anderson
and T. Balch. Atlanta: Georgian Institute of Technology, 2003,139-145.

\bibitem{RGSRM}
Rilley, J. R.; Greggers, U.; Smith, A.D.; Reynolds, D.R.; Menzel,
R. The flight paths of honeybees recruited by the waggle
dance. Nature 2005, 435, 205-207.

\bibitem{Ry}
Ryabko, B. Methods of analysis of animal communi-cation
systems based on the information theory. In: Wiese, K., Gribakin,
F.G., Popov, A.V.,  Renninger, G. (Eds.): Sensory systems of
arthropods. Birkhauser Verlag, Basel, 1993, 627-634.

\bibitem{RR96}
Ryabko, B.; Reznikova, Zh. Using Shannon Entropy and Kolmogorov
Complexity to study the communicative system and cognitive
capacities in ants. Complexity 1996, 2, 37-42.

\bibitem{SST}
Savage-Rumbaugh, E. S.; Shanker, S. G.; Taylor, T. J. Apes,
language and the human mind. Oxford: Oxford University Press, 1998.

\bibitem{SC}
Seyfarth, R. M.; Cheney, D. L. The assessment by vervet
monkeys of their own and another species' alarm calls. Animal
Behaviour 1990, 40, 754-764.

\bibitem{S}
Shannon, C. E. A mathematical theory of communication,  Bell Sys.
Tech. J., vol. 27, pp. 379-423 and pp. 623-656, 1948.


\bibitem{Wa}
Wasmann, E. Die psychischen Fahigkeiten der Ameisen.
Zoologica 1899, 26, 1-133









\end{thebibliography}
\makeatletter
\renewcommand\@biblabel[1]{#1. }
\makeatother

\end{document}